\documentclass[sigconf]{acmart}
\usepackage{amsmath}
\usepackage{amssymb}
\usepackage{booktabs}
\usepackage{color}
\usepackage{color}
\usepackage[export]{adjustbox}
\usepackage{float}
\usepackage{graphicx}
\usepackage{ifpdf}
\usepackage{ifpdf}
\usepackage[utf8]{inputenc}
\usepackage{keyval}  
\usepackage{listings}   
\usepackage{moresize}
\usepackage{multirow}
\usepackage{multirow}
\usepackage{paralist}    
\usepackage{rotating}
\usepackage{soul}
\usepackage{subfigure}
\usepackage{url}
\usepackage{wrapfig}    
\usepackage{xspace}
\usepackage{srcltx}

\definecolor{listinggray}{gray}{0.95}
\definecolor{darkgray}{gray}{0.7}
\definecolor{commentgreen}{rgb}{0, 0.4, 0}
\definecolor{darkblue}{rgb}{0, 0, 0.4}
\definecolor{middleblue}{rgb}{0, 0, 0.7}
\definecolor{darkred}{rgb}{0.4, 0, 0}
\definecolor{brown}{rgb}{0.5, 0.5, 0}
\definecolor{dkgreen}{rgb}{0,0.5,0}
\definecolor{orange}{rgb}{1,.5,0}
\definecolor{dandelion}{cmyk}{0,0.29,0.84,0}

\newif\ifdraft
\ifdraft
 \newcommand{\N}[1]{\textbf{NOTE: #1}\xspace}
 \newcommand{\jhanote}[1]{ {\textcolor{red} { ***SJ: #1 }}}
 \newcommand{\amnote}[1]{ {\textcolor{dkgreen} { ***AM #1 }}} 
 \newcommand{\mtnote}[1]{ {\textcolor{orange} { ***MT: #1 }}}
 \newcommand{\tnnote}[1]{ {\textcolor{orange} { ***TN: #1 }}}
 \newcommand{\wenote}[1]{ {\textcolor{blue} { ***WE: #1 }}}
 \else
 \newcommand{\N}[1]{}
 \newcommand{\jhanote}[1]{}
 \newcommand{\amnote}[1]{}
 \newcommand{\mtnote}[1]{}
 \newcommand{\tnnote}[1]{}
 \newcommand{\wenote}[1]{}
\fi

\newcommand{\B}[1]{\textbf{#1}\xspace}

\newcommand{\I}[1]{\textit{#1}\xspace}

\lstdefinestyle{myListing}{
  frame=single,   
  backgroundcolor=\color{listinggray},  
  language=C,       
  basicstyle=\ttfamily \footnotesize,
  breakautoindent=true,
  breaklines=true
  tabsize=2,
  captionpos=b,  
  aboveskip=0em,
  belowskip=-2em,
}      

\lstdefinestyle{myPythonListing}{
  frame=single,   
  backgroundcolor=\color{listinggray},  
  language=Python,       
  basicstyle=\ttfamily \scriptsize,
  breakautoindent=true,
  breaklines=true
  tabsize=2,
  captionpos=b,  
}

\ifpdf
\DeclareGraphicsExtensions{.pdf, .jpg, .tif}
\else
\DeclareGraphicsExtensions{.ps,  .eps, .jpg}
\fi

\begin{document}

\title{Characterizing the Performance of Executing Many-tasks on Summit}

\author{Matteo Turilli}
\affiliation{Rutgers University}
\email{matteo.turilli@rutgers.com}
\authornote{Co‐first author}
\author{Andre Merzky}
\affiliation{Rutgers University}
\email{andre@merzky.net}
\authornotemark[1]
\author{Thomas Naughton}
\affiliation{Oak Ridge National Laboratory}
\email{naughtont@ornl.gov}
\authornote{This manuscript has been authored by UT-Battelle, LLC under
		Contract No. DE-AC05-00OR22725 with the U.S. Department of Energy.
		The United States Government retains and the publisher, by accepting
		the article for publication, acknowledges that the United States
		Government retains a non-exclusive, paid-up, irrevocable, worldwide
		license to publish or reproduce the published form of this
		manuscript, or allow others to do so, for United States Government
		purposes. The Department of Energy will provide public access to
		these results of federally sponsored research in accordance with the
		DOE Public Access Plan
		(\url{http://energy.gov/downloads/doe-public-access-plan}).}
\author{Wael Elwasif}
\affiliation{Oak Ridge National Laboratory}
\email{elwasifwr@ornl.gov}
\authornotemark[2]
\author{Shantenu Jha}
\affiliation{Rutgers University}
\affiliation{Brookhaven National Laboratory}
\email{shantenu.jha@rutgers.edu}

\begin{abstract}
%!TEX root = prrte-jsrun.tex

Many scientific workloads are comprised of many tasks, where each task is an independent simulation or analysis of data. The execution of millions of tasks on heterogeneous HPC platforms requires scalable dynamic resource management and multi-level scheduling. RADICAL-Pilot  (RP) -- an implementation of the Pilot abstraction, addresses these challenges and serves as an effective runtime system to execute workloads comprised of many tasks. In this paper, we characterize the performance of executing many tasks using RP when interfaced with JSM and PRRTE on Summit: RP is responsible for resource management and task scheduling on acquired resource; JSM or PRRTE enact the placement of launching of scheduled tasks. Our experiments provide lower bounds on the performance of RP when integrated with JSM and PRRTE. Specifically, for workloads comprised of homogeneous single-core, 15 minutes-long tasks we find that: PRRTE scales better than JSM for > O(1000) tasks; PRRTE overheads are negligible; and PRRTE supports optimizations that lower the impact of overheads and enable resource utilization of 63\% when executing O(16K), 1-core tasks over 404 compute nodes.
% This work provides the foundations
% upon which heterogeneous workloads can be executed at scale on Summit.

% PRRTE enables better scalability
% than JSM when executing homogeneous many-task applications at scales larger
% than 987 concurrent task executions; (2) up to the scale currently supported,
% PRRTE individual overheads are negligible when compared to other overheads;
% and (3) PRRTE open source code enables optimizations that lower the impact of
% aggregated overheads over the execution of the considered workloads. 

% are executed on up to 410 compute
% nodes of Summit. Further, based on the insight gained by our
% characterization, we showed the results of our optimization when executing
% 16384 18-core, 15 minutes-long tasks on up to 3511 compute nodes, i.e., 76\%
% of the resources available on Summit.

\end{abstract}

\maketitle

% ---------------------------------------------------------------------------
% 1. INTRODUCTION
% ---------------------------------------------------------------------------
\section{Introduction}\label{sec:intro}

%!TEX root = prrte-jsrun.tex

% 1. Motivate many task workloads. 
% Outline challenges. Include node heterogeneity.
% 2. What is a Task? Examples of many tasks workloads

% 3. Pilot systems execute many tasks and provide an effective model for
% dynamic resource management 

% 4. However they rely on integration with system software to 
% deliver performance and scalability
% Here we investigate RP --- which is used to support....
% on Summit and perform a comparitive
% investigate of its use of JSRUN and PRRTE.

Advances in high-performance computing (HPC) have traditionally focused on
scale, performance and optimization of applications with a large but single
task. Workloads of many scientific applications however, are comprised of many
tasks --- where each task is an independent computing unit running on one or
more nodes, which must be collectively executed and analyzed.

As an example, \textit{ensemble-based} computational methods have been
developed to address limitations in single molecular dynamics simulations,
where parallelization is limited to speeding up each individual, serialized,
time step\cite{cheatham-cise15}. Unlike traditional high-throughput
``embarrassingly'' parallel workloads, these workloads require modest task
coordination and inter-task communication (though very infrequent relative to
intra-task communication). Multiple simulation tasks are executed
concurrently, and various physical or statistical principles are used to
combine the output of these tasks, often iteratively and asynchronously. Tens
to hundreds of thousands of such tasks are currently needed to adequately
sample or investigate the physical phenomenon of interest. Proper sensitivity
analysis and uncertainty quantification can increase the total number of tasks
by several orders of magnitude.

The execution of millions of tasks on modern HPC platforms is faced with many
challenges.  A tension exists between the workload's resource utilization
requirements and the capabilities of traditional HPC system software. It
requires flexible and dynamic resource management of heterogeneous many-core
nodes.

The Pilot abstraction decouples workload specification, resource management,
and task execution. 
% via job placeholders and late-binding of tasks to resources. 
Pilot systems -- implementations of the Pilot abstraction, submit job
placeholders (i.e. pilots) to the resource scheduler. Once active, the pilot
accepts and executes tasks directly submitted to it. Tasks are thus executed
within time and space boundaries set by the resource scheduler. % Pilot
% systems address the aforementioned tension by accessing HPC resources via
% their centralized schedulers while independently scheduling tasks on acquired
% resources.
By implementing multi-level scheduling and late-binding, Pilot systems lower
task scheduling overhead, enable higher task execution throughput, and allow
greater control over the resources acquired to execute workloads. The pilot
must interact with and is dependent on system software to manage the task
execution.

RADICAL-Pilot (RP) is a Pilot system that implements the pilot paradigm as
outlined in Ref.~\cite{merzky2018using,turilli2018comprehensive}. RP is
implemented in Python and provides a well defined API and usage modes, and is
being used by applications drawn from diverse domains, from earth sciences and
biomolecular sciences to high-energy physics. RP directly supports their use
of supercomputers or is used as a runtime system by third party workflow or
workload management systems~\cite{turilli2019middleware}.

In this paper, we characterize the performance of executing many tasks using
RP % task execution of RP 
when it is interfaced with JSM and PRTTE on Summit -- a DOE leadership class
machine and currently the top ranked supercomputer on the Top 500 list. Summit
has 4,608 nodes IBM POWER9 processors and each node has 6 NVIDIA Volta V100s,
with a theoretical peak performance of approximately 200 petaFLOPS. JSM is
part of LSF and provides services for starting tasks on compute resources;
PRTTE provides the server-side capabilities for a reference implementation of
the process management interface for ExaScale (PMIx). Specifically, we
describe and investigate the baseline performance of the integration of
RP~\cite{merzky2018using} with JSM and PRRTE. We experimentally characterize
the task execution rates, various overheads, and resource utilization rates.

% (strong) when using these JSRUN and PRRTE.. We characterize the baseline
% performance of the integration between RADICAL-Pilot and PRRTE, and
% RADICAL-Pilot and JSM\@.

% interacts and utilizes 

% which enable the execution of applications with thousands of tasks on Summit.
% Resources for the execution of the application are acquired via Summit`s batch
% system but application tasks are executed via a dedicated scheduler and
% communication and coordination overlay, without passing through Summit's batch
% system.

% ---------------------------------------------------------------------------
% 3. BACKGROUND
% ---------------------------------------------------------------------------
\section{Background}\label{sec:back}

We characterize the performance of integrating RADICAL-Pilot (RP) and PMIx
Reference RunTime Environment (PRRTE), and RP and IBM Job Step Manager
(JSM)\@. These enable the concurrent execution of thousands of application
tasks on Summit.

% Resources for the execution of the application are acquired via Summit's
% batch system but application tasks are executed via a dedicated scheduler
% without passing through Summit's batch system.

% -------------------------------------------------------------------------
\subsection{Process Management Interface for Exascale}\label{ssec:pmix}

The Process Management Interface for Exascale~(PMIx)~\cite{www:pmix-standard}
is an open source standard that extends the prior PMI~v1~\&~v2 interfaces
used to launch tasks on compute resources.  PMIx provides a method for tools
and applications to interact with system-level resource mangers and process
launch mechanisms.  PMIx provides a bridge between such clients and 
underlying execution services, e.g., process launch, signaling, event
notification.  The clients communicate with PMIx enabled servers, which may
support different versions of the standard.  PMIx can also be used as a
coordination and resource discovery mechanism, e.g., machine topology
information.

%PMIx provides a standard method for resource provides (e.g., batch job
%managers, runtimes) to expose capabilities to applications and tools.

% -------------------------------------------------------------------------
\subsection{IBM Job Step Manager}\label{ssec:jsrun} % (JSM)

The IBM Spectrum Load Sharing Facility~(LSF) software stack is used to manage
the resources for the Summit system.  This includes a job scheduler that
manages resources and provides allocations based on user provided
submissions. The Job Step Manager~(JSM) provides services for starting tasks
on compute resources within an allocation~\cite{www:jsm}. The \texttt{jsrun}
command enables a user to launch an executable on the remote nodes within a
user's job allocation.

%When a user runs a command using JSM, a daemon is started (\texttt{jsmd}) on
%the nodes and is responsible for launching the user's processes. 

When the user is allocated a collection of compute nodes by the batch system,
a daemon (\texttt{jsmd}) is launched on each of the compute nodes in the
allocations. These daemons are then responsible for launching the user's
processes on their respective nodes in response to future \texttt{jsrun}
commands. There are two startup modes for launching the \texttt{jsmd}
daemons: \emph{SSH mode} and \emph{non-SSH mode}~\cite{www:jsm}. As the name
suggests, when running in \emph{SSH mode}, Secure Shell is used for launching
the the \texttt{jsmd} processes on the remote nodes of the allocation. The
other option leverages the IBM Cluster Systems Manager~(CSM) infrastructure
to bootstrap the JSM daemons within the allocation. Currently, the default
mode on Summit is to use CSM\@. Once the JSM daemons are launched, internally
the daemons use PMIx~\cite{www:pmix-standard} to launch, signal, and manage
processes on the compute resources.

%\wenote{I think jsmd is launched as part of the allocation, not as part of
%each jsrun invocation. ssh into a newly allocated node shows the following
%process {\tt/opt/ibm/spectrum\_mpi/jsm\_pmix/bin/admin/jsmd\ csm=562789
%batch2 --peer -ptsargs -p 10.134.0.42,56331 -r 1024,16384 -t 60 }}
%\tnnote{You may be correct, I was using docs, but maybe summit is now setup
%differently.  Is there value in adding that the jsmd is started a the user?
%Maybe just eliminate this detail entirely?}
%\tnnote{I removed qualifcation on UID/owner of jsmd task}

% \tnnote{SHOULD WE ADD that we used JSM/jsrun 10.03.00, which was built with
% PMIx 3.1.3rc1}

% \wenote{Maybe later when we describe the experiment}

% batch4: $ jsrun --version jsrun (Job Step Manager) 10.03.00.01rtm3 [Jun 11,
% 2019] built with PMIx 3.1.3rc1
% https://www.ibm.com/support/knowledgecenter/SSWRJV_10.1.0/jsm/10.3/base/jsm_kickoff.html

% -------------------------------------------------------------------------
\subsection{PMIx Reference RunTime Environment}\label{ssec:prrte} % (PRRTE)

% The Process Management Interface for
% Exascale~(PMIx)~\cite{www:pmix-standard} provides a standard for tools and
% applications to interact with system-level resource managers and launchers.
% This PMIx layer provides a bridge between clients and the underlying
% runtime services, e.g., process launch and event notification.

A reference implementation of the PMIx server-side capabilities is available
via the PMIx Reference RunTime Environment~(PRRTE)~\cite{castain:pc18:pmix}.
PRRTE leverages the modular component architecture~(MCA) that was developed
for Open~MPI~\cite{gabriel04:_open_mpi}, which enables execution time
customization of the runtime capabilities.  The PRRTE implementation provides
a portable runtime layer that users can leverage to launch a PMIx server.

PRRTE includes a persistent mode called Distributed Virtual Machine~(DVM),
which uses system-native launch mechanisms to bootstrap an overlay runtime
environment that can then be used to launch tasks via the PMIx interface.
This removes the need to bootstrap the runtime layer on each invocation for
task launch.  Instead, after the launch of the DVM, a tool connects and sends
a request to start a task. The task is processed and then generates a launch
message that is sent to the PRRTE daemons. These daemons then launch the
task. Internally, this task tracking is referred to as a \emph{PRRTE job},
not to be confused with the batch job managed by the system-wide scheduler.
The stages of each PRRTE job are tracked from initialization through
completion.

We can roughly divide the lifetime of a PRRTE job into the following stages
(marked by internal PRRTE state change events): (i) \texttt{init\_complete}
to \texttt{pending\_app\_launch}---time to setup the task and prepare launch
details; (ii) \texttt{sending\_launch\_msg} to \texttt{running}---time to
send the process launch request to PRRTE daemons and to enact them on the
target nodes; and (iii) \texttt{running} to
\texttt{notify\_complete}---duration of the application plus time to collect
task completion notification.

In our experiments, we record the time for the transition between these
stages to provide insights on the time spent in the runtime layer when
running tasks driven by RP. It should be noted that these phases
do not include time between the user launching a PRRTE task and PRRTE
initiating processing for this task (e.g., due to file system delays or
dynamic libraries loading).

% \tnnote{SHOULD WE ADD that we used PRRTE v1.0.0 with 2 minor patches and
% PMIx 3.1.3}

%%%%%
% PRRTE JOB STATES
% ----------------
% PENDING INIT
% # the jobid is not set for 'PENDING INIT' so no way to track/match this point
% INIT_COMPLETE
% # first marker for jobid is the 'INIT_COMPLETE' stage
% PENDING ALLOCATION
% ALLOCATION COMPLETE
% PENDING DAEMON LAUNCH
% ALL DAEMONS REPORTED
% VM READY
% PENDING MAPPING
% MAP COMPLETE
% PENDING FINAL SYSTEM PREP
% PENDING APP LAUNCH
% SENDING LAUNCH MSG
% RUNNING
% #
% # app running
% #
% NORMALLY TERMINATED
% NOTIFY COMPLETED
% # parent watcher also does 'NORMALLY TERMINATED' and 'NOTIFY COMPLETED'
% # so the technically the final marker for a given job is the time of
% # the parent of our job reporting 'NOTIFY COMPLETE'
%
%
% - 'INIT_COMPLETE' to 'PENDING APP LAUNCH' (or 'SENDING LAUNCH MSG')
% - 'SENDING LAUNCH MSG' to 'RUNNING'
% - 'RUNNING' to 'NORMALLY TERMINATED' (duration of the app + delta)
% - 'RUNNING' to (parent) 'NOTIFY COMPLETED' (duration of app + full delta)
%%%%%

% -------------------------------------------------------------------------
\subsection{RADICAL-Pilot}\label{ssec:rp}

\begin{figure}
  \centering
  \includegraphics[trim=0 0 0 0,clip,width=0.45\textwidth]{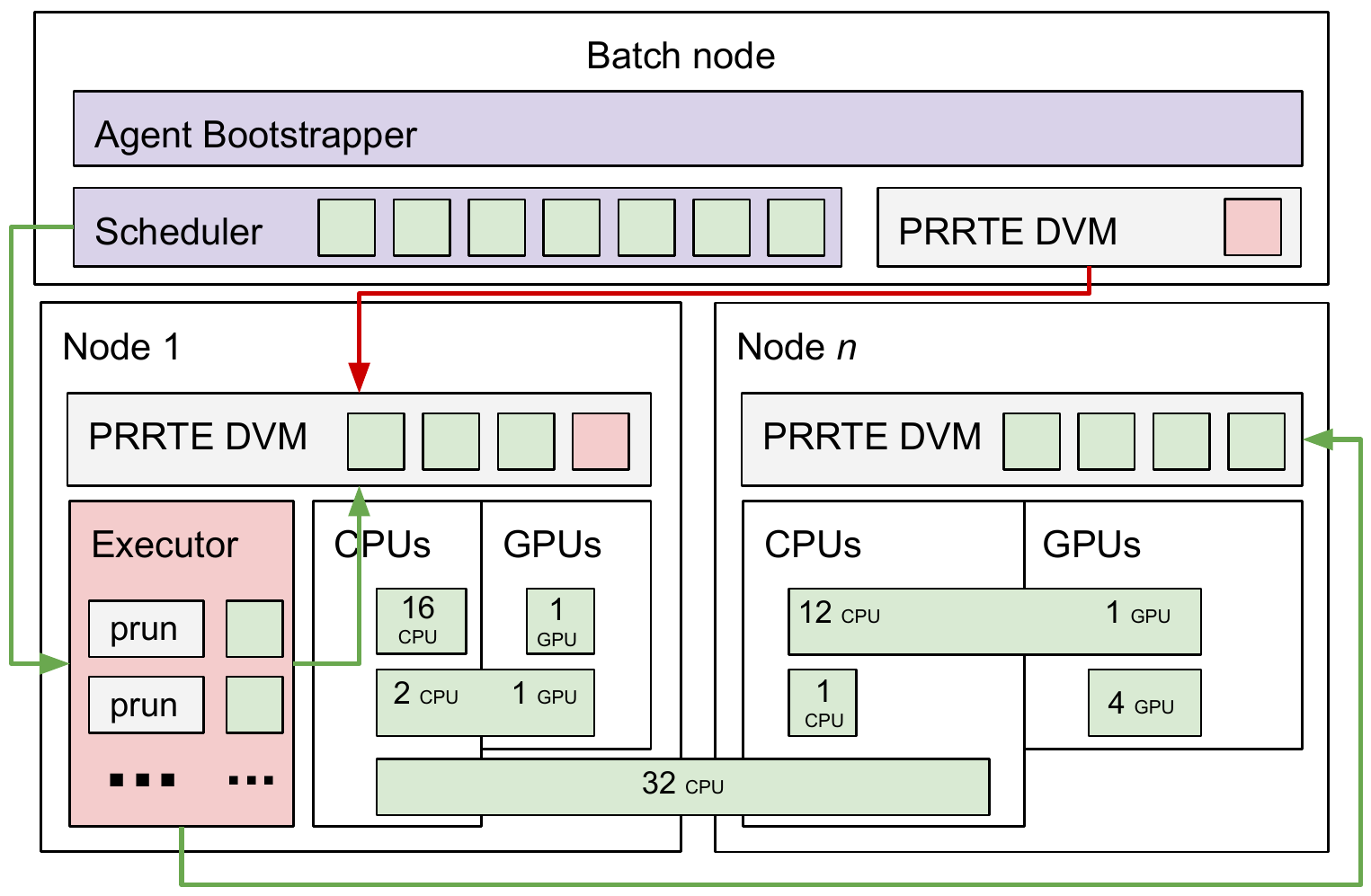}
  \Description{An architecture block diagram describing the integration
               between RP and PRRTE.}
    \caption{\footnotesize Deployment of RP on Summit with
           PRRTE/DVM.}\label{fig:rp-on-summit}
\end{figure}

RP~\cite{merzky2018using} is a runtime system designed to decouple resource
acquisition from task execution. As every pilot system, RP acquires resources
by submitting a batch job, then bootstraps dedicated software components on
those resources to schedule, place and launch application tasks, independent
from the machine batch system~\cite{turilli2018comprehensive}. Scheduling,
placing and launching capabilities are specific to each HPC platform, which
makes supporting diverse platforms with the same code base challenging. RP
can execute single or multi core/GPU tasks within a single compute node, or
across multiple nodes. RP isolates the execution of each tasks into a
dedicated process, enabling concurrent and sequential execution of
heterogeneous tasks by design.

RP is a distributed system designed to instantiate its components across
available resources, depending on the platform specifics. Each components can
be individually configured so as to enable further tailoring while minimizing
code refactoring. RP uses RADICAL-SAGA~\cite{saga-x} to support all the major
batch systems, including Slurm, PBSPro, Torque and LSF\@. RP also supports
many methods to perform node and core/GPU placement, process pinning and task
launching like, for example, aprun, JSM, PRRTE, mpirun, mpiexec and ssh.

RP is composed of two main components: Client and Agent. Client executes on
any machine while Agent bootstraps on one of Summit's batch nodes. Agent is
launched by a batch job submitted to Summit's LSF batch system via
RADICAL-SAGA\@. After bootstrapping, Agent pulls bundles of tasks from
Client, manages the tasks' data dependences if any, and then schedules tasks
for execution via either JSM/LSF or PRRTE/DVM\@.

How Agent deploys on Summit depends on several configurable parameters like,
for example, number of sub-agents, number of schedulers and executors per
sub-agent, method of communication between agent and sub-agents, and method
of placing and launching tasks for each executor of each sub-agent. A default
deployment of Agent instantiates a single sub-agent, scheduler and executor
on a batch node of Summit. The executor calls one \texttt{jsrun} command for
each task, and each \texttt{jsrun} uses the JSMD demon to place and launch
the task on work nodes resources (thread, core, GPU).

Fig.~\ref{fig:rp-on-summit} shows the deployment of RP/PRRTE\@ Agent on a
batch node and one sub-agent on a compute node. In this configuration, RP
uses SSH to launch sub-agents on compute nodes and then PRRTE/DVM to place
and launch tasks across compute nodes. This configuration enables the
sub-agent to use more resources and, as shown in the next section, improves
scalability and performance of task execution. Note that, independent from
the configuration and methods used, RP can concurrently place and launch
different types of tasks that use different amount and types of resources.

% ---------------------------------------------------------------------------
% 4. PERFORMANCE CHARACTERIZATION
% ---------------------------------------------------------------------------
\section{Performance Characterization}\label{sec:performance}

The performance space of RP is vast, including the execution of both
homogeneous and heterogeneous tasks, with and without data dependences and at
both small and large scales. We thus divide our performance characterization
in three phases: (1) scaling the concurrent execution of short, single-core
tasks with both JSM and PRRTE\@; (2) optimizing baseline performance for
homogeneous real-life workloads with the best performing between JSM and
PRRTE\@; (3) tailoring performance to data-intensive, compute-intensive and
GPU-intensive workloads. We present the results of the first phase, offering
a baseline that we use to drive our development process.

Task here indicates a self-contained executable, executed as one or more
processes on the operating system of a Summit compute node. RP, JSM and PRRTE
introduce time overheads in tasks execution. These systems require time to
schedule, place and launch the task executions. We quantify and compare these
overheads, measuring how they change with the scaling of the number of
concurrently executed tasks and the amount of used resources.

We differentiate between individual overheads and the integration of these
overheads over the execution of the workload. Individual overheads account
for the amount of time that single operations add to the execution time of a
task. For example, how much time RP takes to schedule a task or PRRTE takes
communicating to launch that task. Aggregated overheads indicate how much
time performing a group of operations adds to the execution of all the
workload tasks. Aggregated overheads account for the overlapping of multiple
concurrent operations. For example, given 10 tasks, a scheduling rate of 1
task/s and a scheduling time of 5s for task, the aggregated scheduling
overhead would be 15s for full concurrency, while the individual scheduling
overhead for each task would be 5s.

The aggregation of the individual overheads across the entire execution
determines how available resources are utilized when executing the workload.
RP, JSM and PRRTE require resources to perform their operations and some of
these operations may partially or globally block the use of available
resources. We measure resource utilization showing the portion of resources
used or blocked by each system and the percentage of available resources
utilized to execute the workload.

%----------------------------------------------------------------------------
\subsection{Experiments Design}\label{ssec:exp_design}

We perform 4 experiments to measure and compare the performance of RP, JSM
and PRRTE when concurrently executing many-tasks workloads on Summit. Task
execution requires assigning suitable resources to the tasks, placing them on
resources (i.e., a specific compute node, core, GPU or hardware thread) and
then launching the execution of those tasks. RP tracks both tasks and
available resources, scheduling the former onto the latter; JSM or PRRTE
enact task placement and launching.

\begin{figure}
  \centering
    \includegraphics[trim=0 0 0 80,clip,width=0.49\textwidth,valign=t]{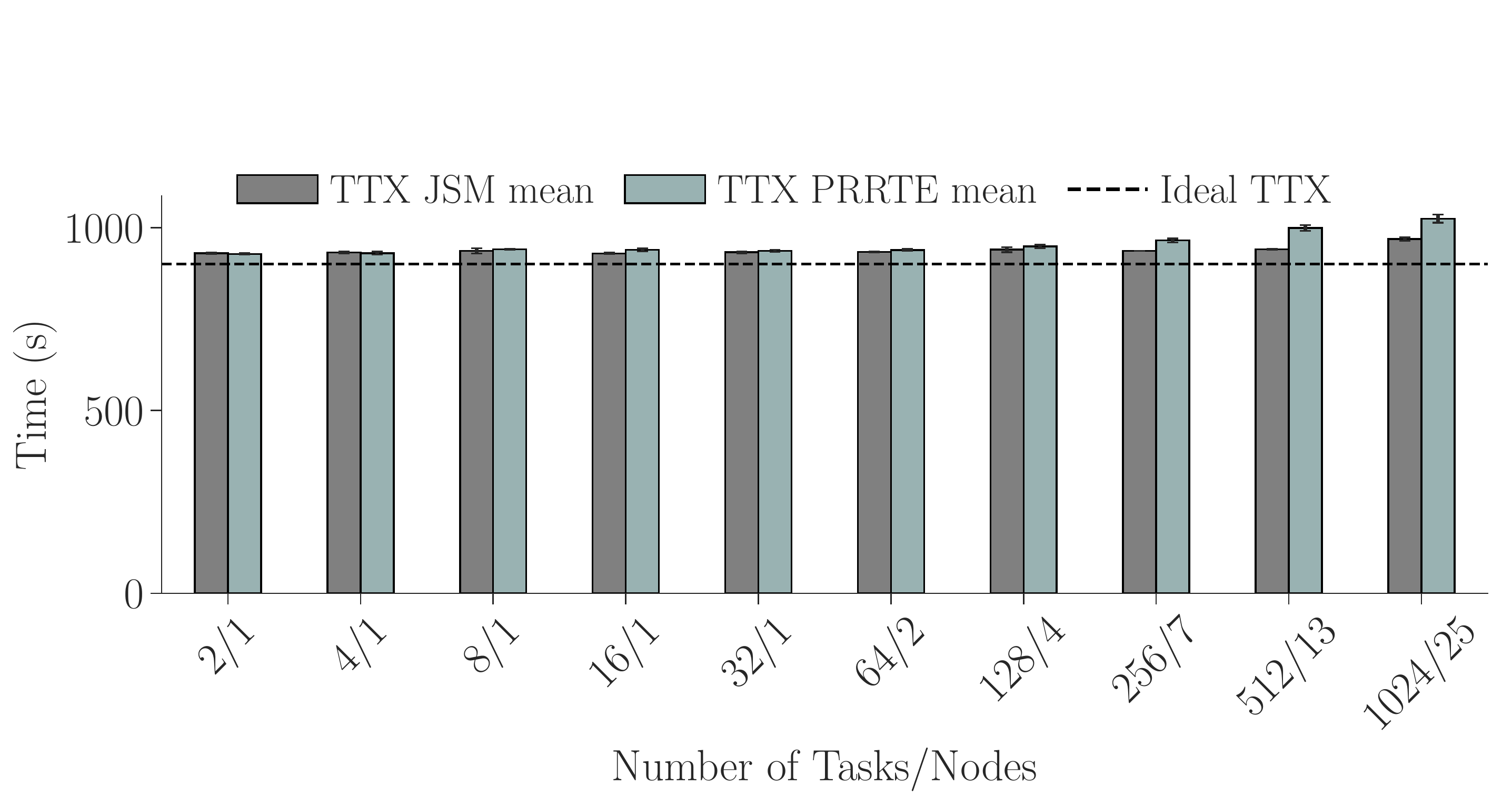}
    \includegraphics[trim=0 0 0 80,clip,width=0.49\textwidth,valign=t]{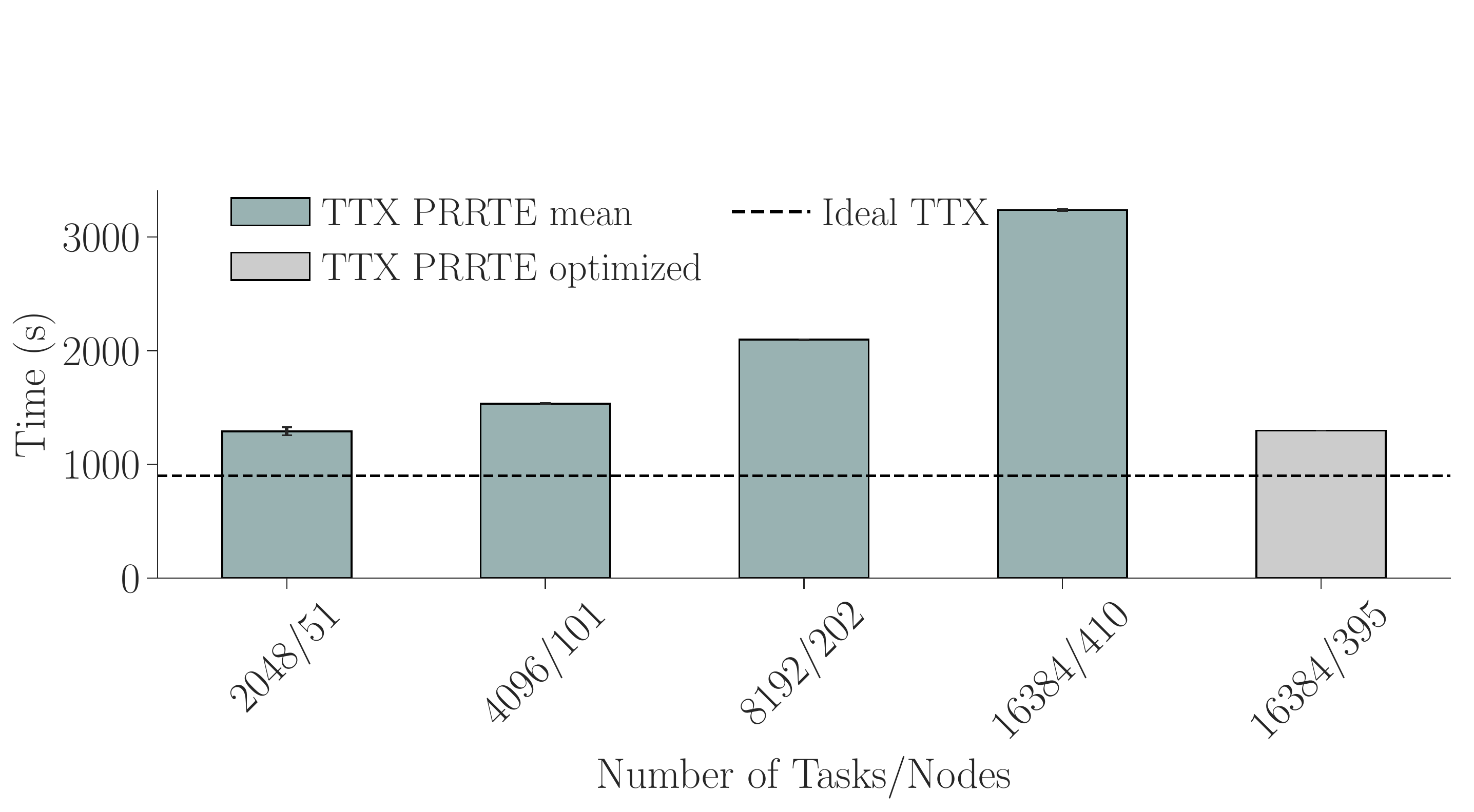}
  \Description{Bar plot.}
  \caption{\footnotesize Measured and ideal total execution time (TTX) of the
           workloads of Experiment 1--3 (green) and 4
           (gray).}\label{fig:exp-ttc}
  \vspace*{-1.7em}
\end{figure}

Experiment 1 quantifies the aggregated overhead of RP measured as the time
required to acquire the workload and scheduling its tasks on the available
resources. Experiment 1 measures this overhead with both JSM and PRRTE\@.
Experiment 2 quantifies the aggregated overhead of JSM and PRRTE measured as
the time from when they receive the tasks from RP to when the tasks start to
execute. Experiment 3 measures RP and PRRTE aggregated overheads and resource
utilization for scales beyond those currently supported by JSM\@. Experiment
4 shows the performance improvement obtained by reducing the overheads
measured in Experiments 1--3.

In Experiment 1--4 we execute workloads with between 2 and 16384 tasks, each
requiring 1 core and executing for 900s. Given Summit architecture, we
utilize between 1 and 410 compute nodes, i.e., 42 to 17220 cores, using the
SMT1 setting for Summit nodes. Our experiments maximize execution concurrency
and therefore the pressure on RP, JSM and PRRTE capabilities. Any lower
degree of execution concurrency would require less scheduling, placements and
executions, resulting in a better performance of our systems. As such, our
experiments measure the baseline of the combined scaling performance of RP,
JSM or PRRTE on Summit for homogeneous compute-intensive, multi-tasks
workloads.

Experiment 1--4 make a set of reasonable simplifications: each task executes
the \texttt{stress} command for exactly 900s, a trade off between core
allocation cost and the need to stress RP, JSM and PRRTE performance. We do
not perform I/O operations as they would be irrelevant to our
characterization. JSM and PRRTE do not manage task-level data while RP only
links data on the available filesystems and ensures locality of output data.
In this context, data performance depends on the storage systems and the
executable capabilities and should be independently characterized.

Analogously, in our experiment we do not use real workloads executables. RP,
JSM and PRRTE make sure that the executable of a task is launched on the
required resources but play no role on their ongoing execution. The
executable is self-contained and completely independent from RP, JSM and
PRRTE\@. Thus, the measurements we present apply to every homogeneous
workload, independent of the scientific domain, and the specifics of the code
executed.

For our experiments we use JSM/jsrun 10.03.00, built for PMIx 3.1.3rc1; PRRTE
v1.0.0 with 2 minor patches, built for PMIx 3.1.3; and RADICAL-Cybertools
v0.70.3. The data, analysis and code of our experiments is available
at~\cite{turilli2019prrte-exp}.

Fig.~\ref{fig:exp-ttc} shows the total execution time (TTX) of the workloads
of Experiment 1--4. The black line indicates the ideal execution time, when
both software and hardware overheads are assumed to be zero. As expected, the
aggregated overheads of the execution increase with the number of tasks and
compute nodes utilized. The last column shows Experiment 4 and the marked
improvement made by addressing the overheads measured in Experiment 1--3.

%----------------------------------------------------------------------------
\subsection{Experiment 1: RP Aggregated Overhead}\label{ssec:exp1}

\begin{figure}
  \centering
  \includegraphics[trim=0 10 0 0,clip,width=0.49\textwidth]{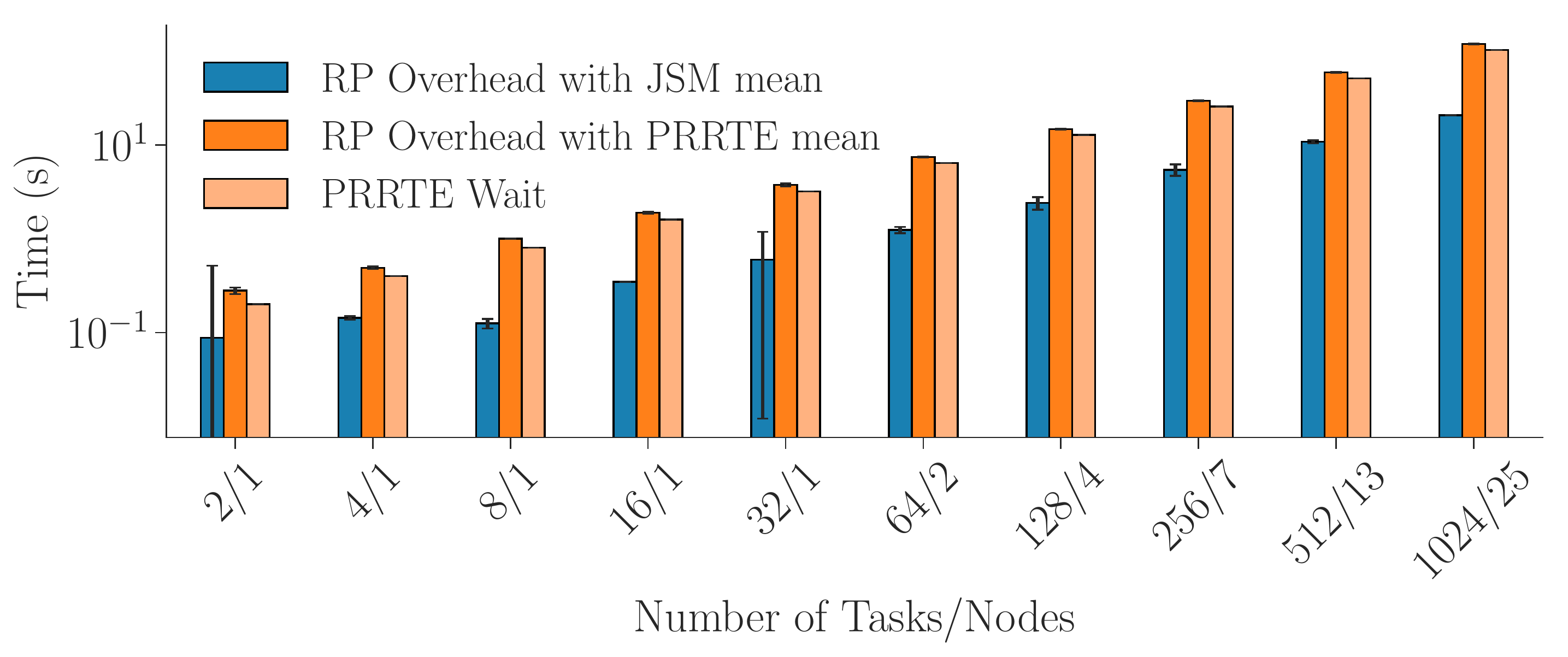}
  \Description{Bar plot.}
  \caption{\footnotesize Experiment 1: RP aggregated
           overheads when scheduling 2--1024 single-core tasks on 1--49
           compute nodes on Summit.}\label{fig:exp1}
\end{figure}

Fig.~\ref{fig:exp1} shows the RP aggregated overhead when using either JSM or
PRRTE to place and launch between 1 and 1024 single-core tasks on between 1
and 49 Summit compute nodes.

The mean of the aggregated overhead of RP grows exponentially with scale but
differences can be noted between JSM and PRRTE when executing 2 to 8 tasks.
With JSM, the aggregated overhead is relatively stable but with large
variability at 2 and 32 tasks. With PRRTE, the aggregated overhead grows with
minimal variability. Assuming ideal concurrency and resource availability,
all the tasks would concurrently execute in 900 seconds. Comparatively, the
mean aggregated overhead of RP is \(<5\%\) of the ideal total execution time
(TTX) with JSM, and \(<25\%\) with PRRTE\@.

Across all scales, the mean of the aggregated overhead of RP is consistently
higher with PRRTE than with JSM\@. This is due to a communication delay we
introduced when pushing tasks to PRRTE\@. RP task scheduling rate is higher
than PRRTE task ingestion rate and exceeding it causes PRRTE to fail. We thus
slow down RP scheduling rate by introducing an artificial 0.1s delay per
task.

PRRTE Wait in Fig.~\ref{fig:exp1} shows the portion of RP aggregated overhead
which is due to the delay we introduced. As we are measuring an aggregated
overhead, the delays accumulates across the whole execution. PRRTE Wait
dominates the RP overhead, showing how, in relative terms, PRRTE overhead is
smaller than the one of JSM\@. Accounting for PRRTE Wait, the mean aggregated
overhead of RP is below 3\% of the ideal execution time with PRRTE\@.

We used test runs to empirically derive the amount of delay to introduce in
RP when communicating with PRRTE\@. Exceeding the submission rate with PRRTE
leads to tasks submission errors that RP could recover at the cost of reduced
utilization.  At a rate of 10 tasks/second we observe stable operation of the
PRRTE DVM\@.

Similar test runs uncovered the failure modes of JSM\@. Originally, exceeding
the sustainable rate of calls to JSM caused the LSF daemon to unrecoverably
fail. This crashed the LSF job with which RP acquired its resources, causing
the failure of the workload execution. Recent updates to LSF on Summit
resolved those issues, and no delays are required for sequential JSM
invocations.

% ---------------------------------------------------------------------------
\subsection{Experiment 2: JSM and PRRTE Aggregated
Overheads}\label{ssec:exp2}

\begin{figure}
  \centering
  \includegraphics[trim=0 13 0 8,clip,width=0.49\textwidth]{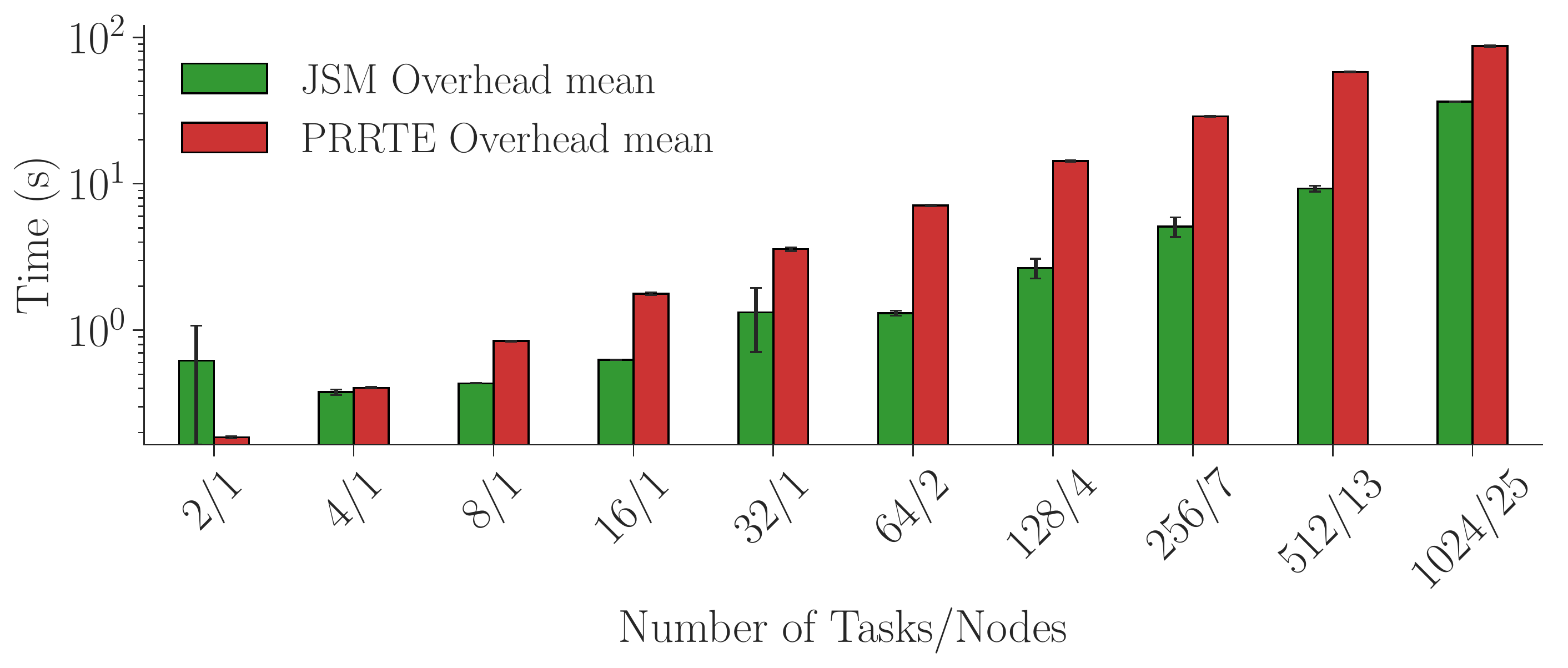}
  \Description{Bar plot.}  
    \caption{\footnotesize Experiment 2: JSM and PRRTE aggregated overheads
           when placing and launching 1--2048 single-core tasks on 1--49
           nodes.}\label{fig:exp2}
\end{figure}

Fig.~\ref{fig:exp2} shows the aggregated overheads of JSM and PRRTE when
executing the same workloads as Experiment 1. Starting from 4/1 tasks/node,
JSM has a smaller aggregated overhead compared to PRRTE\@. PRRTE aggregated
overhead grows exponentially with the number of tasks and nodes while JSM
shows a less well-defined trend across scales. This is because the delay
introduced in RP makes the aggregation of PRRTE individual overheads
additive: the delay is longer than PRRTE task placement and launch time.

The aggregated overheads of RP, JSM and PRRTE do not sum to the total
overhead as measured in Fig.\ref{fig:exp-ttc}. RP can schedule tasks while
other tasks are been assigned and launched by JSM or PRRTE\@. Thus, the
aggregated overhead of RP and those of JSM or PRRTE may overlap during
execution, resulting as a single overhead when projected on TTX\@.

Both aggregated overheads plateau below 1024 tasks and 25 nodes. This is due
to task failure: both JSM and PRRTE can execute up to 967 tasks. Above that,
the remaining tasks fail, creating a steady overhead. This upper bound
depends on the limit to 4096 open files imposed by the system on a batch
node. This results in a maximum of O(1000) tasks as each task consumes at
least three file descriptors: standard input, output, and error files.

% ---------------------------------------------------------------------------
\subsection{Experiment 3: RP and PRRTE at Scale}\label{ssec:exp3}

We overcome the limit of O(1000) concurrent tasks by running multiple
instances of RP executor onto compute nodes and reconfiguring the open files
limit to 65536. This allows up to \({\sim}22000\) concurrent tasks per
executor but this approach works only with PRRTE\@. The open files limits
cannot be increased with JSM, and JSM becomes unstable with concurrent RP
executors. Thus, we could not execute \(>967\) concurrent task with JSM\@.

\begin{figure}

  \centering
  \includegraphics[trim=0 13 0 10,clip,width=0.49\textwidth,valign=t]{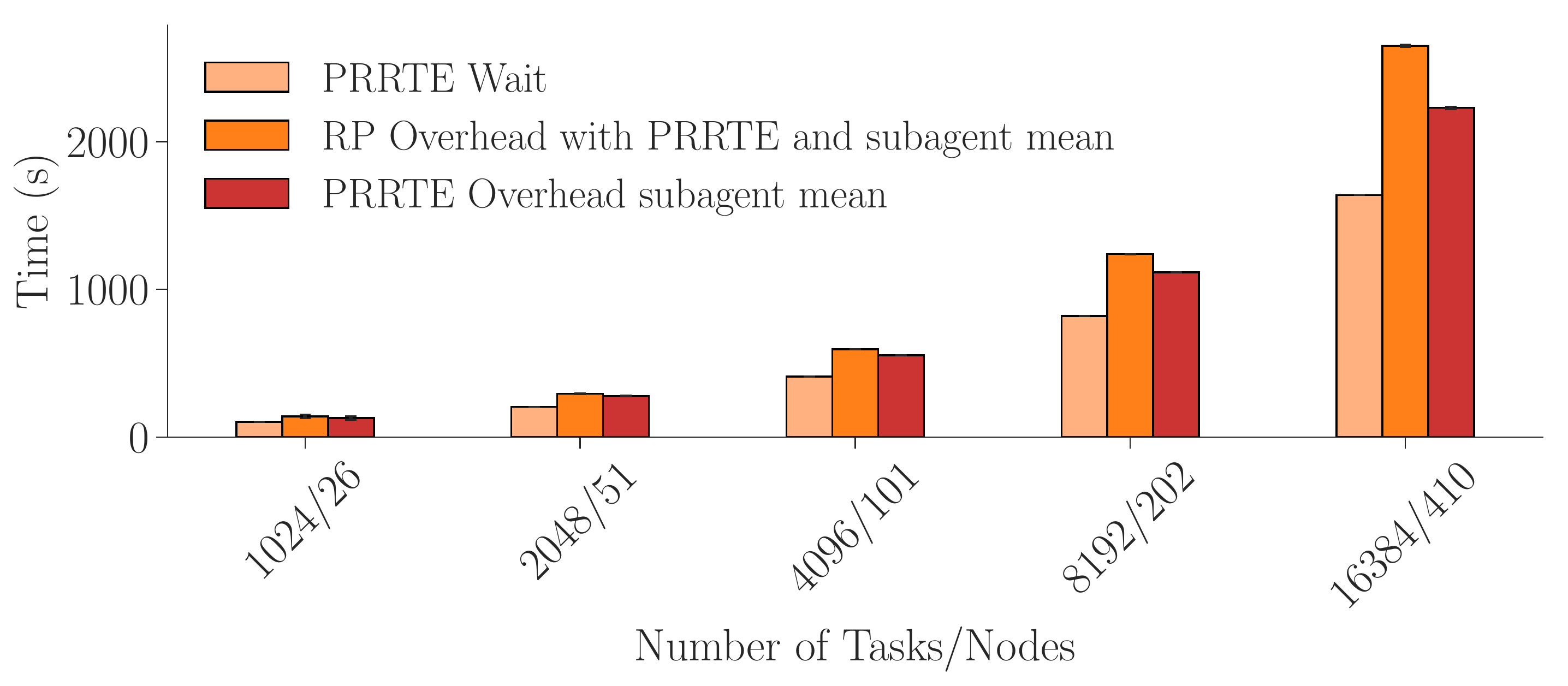}
  \Description{Two bar plots.}
    \caption{\footnotesize Experiments 3: Aggregated overhead of RP and
           of PRRTE when executing 1024--16384 single-core tasks on 26--410
           nodes.}\label{fig:exp3}
\end{figure}

Fig.~\ref{fig:exp3} shows the behavior of the aggregated overheads of both RP
and PRRTE at scale. As already observed in Experiment 2, these overheads grow
exponentially and, for RP, the waiting time introduced when communicating
with PRRTE remains dominant.

16384/401 (single-core) tasks/nodes is the limit currently supported by
RP/PRRTE integration. At 32768/815 tasks/nodes, the DVM of PRRTE crashes,
likely due to the excessive number of communication channels concurrently
opened. Fig.~\ref{fig:exp-ttc} (bottom) shows that the combination of RP and
PRRTE aggregated overhead becomes dominant over TTX at 8192/202 tasks/nodes
so scaling beyond 16384/410 tasks/cores would not be effective.

The aggregated overheads of RP and PRRTE, alongside the TTX of the workload
execution are stable. The variance across runs is minimal, indicating
consistent executions across time. Further, experiments 1--4 executed
\({\sim}200000\) tasks without failure, a measure of the robustness of the
integration between RP and PRRTE and, up to 967 concurrent tasks, of RP and
JSM\@.

Different from JSM and LSF, PRRTE is an open source project. This allows us
to profile the execution of each task inside PRRTE's DVM\@. We collects 40
timestamps, that can be grouped pairwise to provide up to 20 sequential
durations for each task execution. Profiles allow to isolate overheads both
at individual and aggregated levels, enabling to separate in
Fig.~\ref{fig:exp3} (bottom) the aggregated task execution overhead and that
of the DVM\@.

Fig.~\ref{fig:exp3-prrte} (top) shows the breakdown of the aggregated
overhead of PRRTE's DVM for the execution of Experiment 3 workloads. The
dominant aggregated overhead of PRRTE is the time taken to communicate to the
demons on each compute node that a task is ready to execute. As already
noted, this overhead is the sum of each individual overhead as the rate at
which the tasks are queued for execution by RP to PRRTE is too slow to create
any overlapping among communication of initiating the execution of tasks.

Fig.~\ref{fig:exp3-prrte} (bottom) shows the time taken to communicate the
execution of each task within PRRTE's DVM\@. The average time taken by each
individual overhead is 0.034s, and a standard deviation of 0.047s. The
outliers are likely produced by an accumulation in the communication buffer
but over the 16384 tasks, the time taken by this communication is mostly
stable around the mean. For 16384 tasks, the individual overheads sum up to
570s, \({\sim}17\%\) of the TTX of the workload (as shown in
Fig.~\ref{fig:exp-ttc} (bottom)).

% ---------------------------------------------------------------------------
\subsection{Resource Utilization}\label{ssec:utilization}

\begin{figure*}
  \centering
  \includegraphics[trim=0 30 0 80,clip,width=0.99\textwidth]{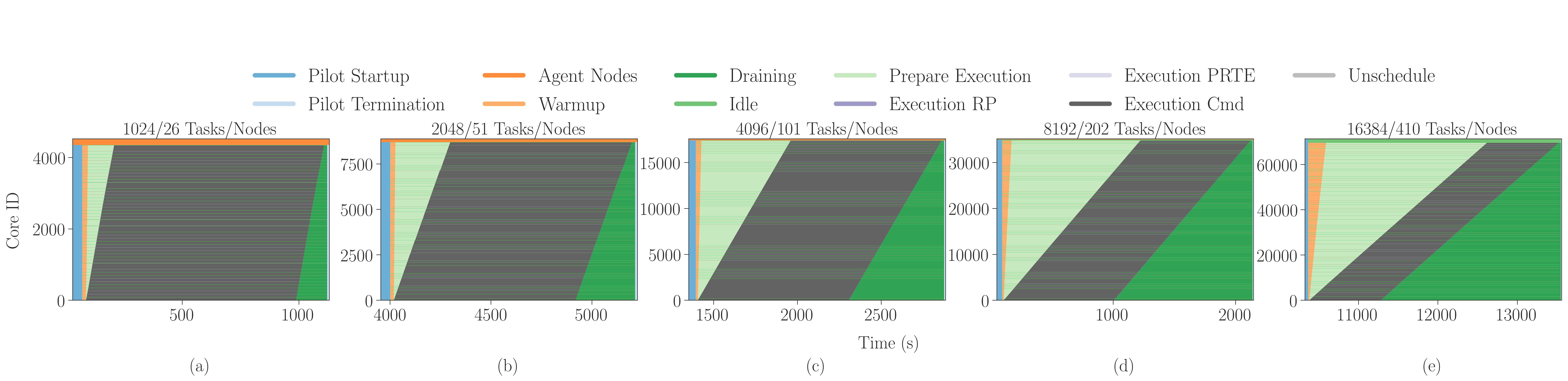}
\Description{Area plot describing RP, PRRTE and workload
             resource utilization}
  \caption{\footnotesize Experiment 3: Resource utilization as the time in
  		   which each available core has been utilized or blocked by RP,
  		   PRRTE or the workload execution.}\label{fig:ru}
\end{figure*}

We measure for how much time each available computing resource is used,
blocked or left idling during the execution of a workload. We focus only on
the runs with RP/PRRTE as relevant behavior is measured only at the largest
scales of our experiments.

Resources become available as soon as Summit's scheduler bootstraps the job
submitted to Summit's batch system on one of the requested compute nodes.
From then on, we distinguish among three main consumers of those resources:
RP, PRRTE and the workload. Each consumer can use the resources to perform
computations, block the resources while a computation is performed, or
resources can idle because they have no active consumers.

The percentage of resource consumed indicate how much of the resources has
been actually used to execute the scientific payload and, in case, where
resources have been wasted while being blocked by other operations.  It is
the most relevant measure of the baseline performance of RP and PRRTE
integration.

Fig.~\ref{fig:ru} shows resource utilization (RU) for Experiment 3. \I{Agent
Nodes} (dark orange) indicates the resources allocated exclusively to RP\@.
\I{Pilot Startup} (blue) shows the time in which the resources are blocked
while RP starts up; and \I{Warmup} (light orange) the time in which resources
are blocked by RP while collecting tasks and scheduling them for execution.
\I{Prepare Execution} (light green) indicates the resources blocked while
waiting to communicate with PRRTE for task execution;
\I{Execution Cmd} (black) marks the time in which tasks use resources for
execution. \I{Draining} (dark green) indicates the resources blocked while
waiting to drain tasks upon their completion; and \I{Pilot Termination}
(light gray) shows the resources blocked while terminating the pilot.

Consistent with the overhead analysis, execution preparation (light green)
and execution draining (dark green) progressively dominate RU with scale.
Execution preparation corresponds to the wait time introduced in RP and
draining time is specular to wait time: the slower is the rate at which tasks
are started, the slower is the rate at which they can be drained. Both
starting and draining time are blocking operations that, at the current
supported launching rates, result in large amount of available resources not
to be used for the execution of the workload.

\I{Execution RP}, \I{Execution PRRTE} and \I{Unschedule} are too small to be
seen when compared to the other measures of RU\@. This indicates that PRRTE
has no appreciable impact over RU during the workload execution. RP impact is
noticeable for exclusive resource allocation (a whole compute node), and for
blocking available resources while bootstrapping and preparing the execution.
\I{Pilot termination} is visible only at the lower scales as it has a mostly
constant impact on RU\@. Fig.~\ref{fig:ru} shows several horizontal green
lines, cutting across each plot, indicating resources idling across the whole
execution. In Experiment 3, these resources are GPUs as our workload does not
use them.

\begin{figure}
  \centering
  \includegraphics[trim=0 0 0 10,clip,width=0.49\textwidth]{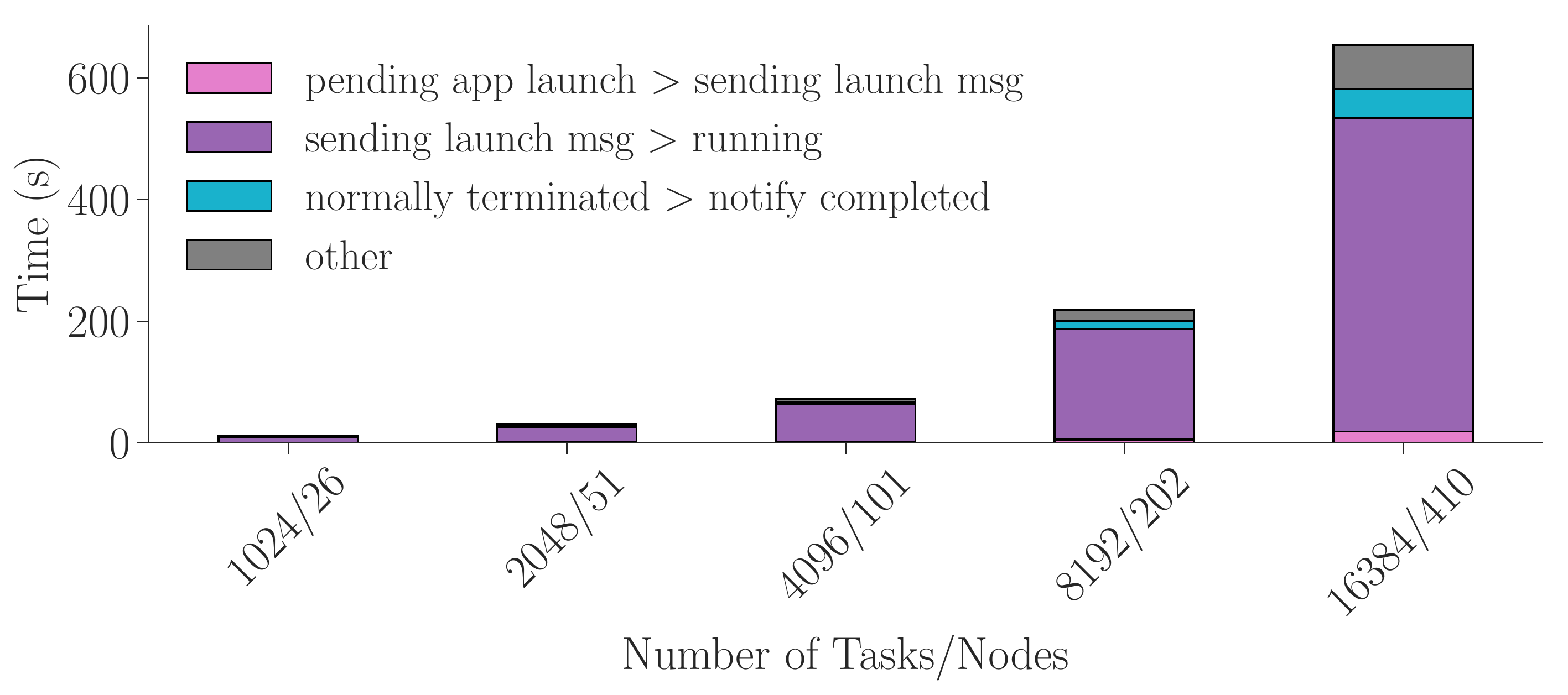}
  \includegraphics[trim=0 10 0 80,clip,width=0.49\textwidth,valign=t]{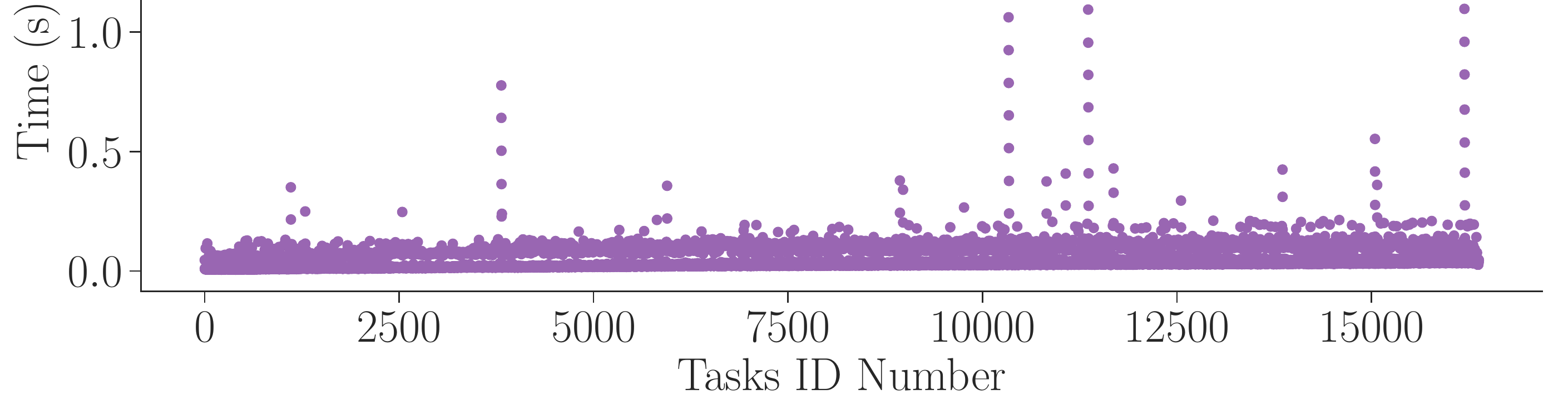}
  \Description{Two bar plots.}
  \caption{\footnotesize Experiment 3: Dominant aggregated overheads of PRRTE
           when executing 1024/26--16834/410 tasks/nodes on
           Summit.}\label{fig:exp3-prrte}
\end{figure}

Table~\ref{tab:ru} details our measures of RU as percentage of the available
resources. Resources used by RP are independent of resource size, thus the
percentage of resource utilized by RP decreases with the size of the pilot.
Similarly, the percentage of available resource blocked while starting the
pilot decreases with scale as startup time is relatively invariant across
pilot sizes. The resources blocked while ``warming up'', i.e., collecting and
scheduling tasks, significantly increases from 8192 tasks onwards. This is
mainly due to scheduling efficiency as RP scheduler performance depends on
the amount of available resources. Consistently with what observed in
Figs.~\ref{fig:exp3-prrte} and~\ref{fig:ru}, PRRTE has a negligible impact on
RU across all the scales.

\begin{table*}
	\caption{\footnotesize Experiments 3--4: Resource utilization (RU)
	expressed as the percentage of resources used of blocked by RP, PRRTE and
	the workload. Last line shows optimized run of Experiment
	4.}\label{tab:ru}
	\centering
	\resizebox{\textwidth}{!}{
	\begin{tabular}{l % p{1.6cm}  % Tasks/Nodes
	                c % p{1.7cm}  % Agent Nodes
	                c % p{1.7cm}  % Pilot Startup
	                c % p{1.1cm}  % Warmup
	                c % p{2.1cm}  % prepare Execution
	                c % p{1.2cm}  % Execution RP
	                c % p{1.8cm}  % Execution PRTE
	                c % p{1.5cm}  % Execution Cmd
	                c % p{1.5cm}  % Unschedule
	                c % p{1.5cm}  % Draining
	                c % p{2.2cm}  % Pilot Termination
	                c % p{0.8cm}  % Idle
				   }
	\toprule
    \B{Tasks / Nodes}     &
    \B{Agent Nodes}       &
    \B{Pilot Startup}     &
    \B{Warmup}            &
    \B{Prep. Execution}   &
    \B{Exec. RP}          &
    \B{Exec. PRRTE}       &
    \B{Exec. Cmd}         &
    \B{Unschedule}        &
    \B{Draining}          &
    \B{Pilot Termination} &
    \B{Idle}              \\
	\midrule
	1024 / 26    & % Tasks/Nodes
	3.846\%      & % Agent Nodes
	3.630\%      & % Pilot Startup
	1.680\%      & % Warmup
	4.510\%      & % Prepare Execution
	0.016\%      & % Execution RP
	0.002\%      & % Execution PRTE
	\B{73.999}\% & % Execution Cmd
	0.001\%      & % Unschedule
	6.149\%      & % Draining
	0.812\%      & % Pilot Termination
	5.355\%   \\  % Idle
	2048 / 51    & % Tasks/Nodes
	1.961\%      & % Agent Nodes
	3.622\%      & % Pilot Startup
	1.603\%      & % Warmup
	9.800\%      & % Prepare Execution
	0.011\%      & % Execution RP
	0.004\%      & % Execution PRTE
	\B{65.313}\% & % Execution Cmd
	0.000\%      & % Unschedule
	11.356\%     & % Draining
	0.867\%      & % Pilot Termination
	5.462\%   \\  % Idle
	4096 / 101   & % Tasks/Nodes
	0.990\%      & % Agent Nodes
	2.698\%      & % Pilot Startup
	1.398\%      & % Warmup
	16.178\%     & % Prepare Execution
	0.013\%      & % Execution RP
	0.002\%      & % Execution PRTE
	\B{54.797}\% & % Execution Cmd
	0.000\%      & % Unschedule
	17.798\%     & % Draining
	0.534\%      & % Pilot Termination
	5.593\%    \\  % Idle
	8192 / 202   & % Tasks/Nodes
	0.495\%      & % Agent Nodes
	2.076\%      & % Pilot Startup
	1.954\%      & % Warmup
	23.375\%     & % Prepare Execution
	0.021\%      & % Execution RP
	0.002\%      & % Execution PRTE
	\B{39.990}\% & % Execution Cmd
	0.001\%      & % Unschedule
	25.570\%     & % Draining
	0.396\%      & % Pilot Termination
	6.120\%    \\  % Idle
	16384 / 410  & % Tasks/Nodes
	0.244\%      & % Agent Nodes
	1.271\%      & % Pilot Startup
	3.309\%      & % Warmup
	28.779\%     & % Prepare Execution
	0.021\%      & % Execution RP
	0.002\%      & % Execution PRTE
	\B{25.596}\% & % Execution Cmd
	0.001\%      & % Unschedule
	32.752\%     & % Draining
	0.256\%      & % Pilot Termination
	7.771\%     \\  % Idle
    \midrule
	16384 / 410  & % Tasks/Nodes
	1.013\%      & % Agent Nodes
	3.265\%      & % Pilot Startup
	6.314\%      & % Warmup
	2.345\%      & % Prepare Execution
	2.421\%      & % Execution RP
	4.988\%      & % Execution PRTE
	\B{63.557}\% & % Execution Cmd
	0.286\%      & % Unschedule
	11.526\%     & % Draining
	0.800\%      & % Pilot Termination
	3.485\%     \\  % Idle
	\bottomrule
	\end{tabular}
	}
\end{table*}

% ---------------------------------------------------------------------------
\subsection{Discussion}\label{ssec:discussion}

Experiments 1--3 and the metrics time to execution (TTX) and resource
utilization (RU) offer three main insight: (1) performance baseline of RP
with JSM or PRRTE on Summit up to the scale currently supported; (2)
characterization of aggregated and individual overheads of RP, JSM and
PRRTE\@; and (3) performance evaluation of RP for TTX and RU\@.

One of the main goals of our performance baseline is to guide the engineering
with which we integrate RP, JSM and PRRTE\@. The analysis we presented shows
that the waiting time between RP and PRRTE communication is the main barrier
to scalability. Figs.~\ref{fig:exp1} and Fig.~\ref{fig:exp2} show that PRRTE
Wait is the dominant aggregated overhead. The analysis of
Fig.~\ref{fig:exp3-prrte} showed how this waiting time determines the
aggregation of PRRTE overheads into the sum of non-overlapping task
overheads. Further, Fig.~\ref{fig:ru}, shows that the waiting time reduces up
to 3/4 the resources that the workload can utilize for execution.

The delay we introduced is conservative so to guarantee no failure in task
execution. In many real life use cases, task failure can be managed with
fault tolerance as done, for example, with RADICAL Ensemble Toolkit (EnTK)
and RP when executing seismic inversion on
Titan~\cite{balasubramanian2018harnessing}. There, we scaled the execution up
to 131,000 cores, resubmitting \({\sim}15\)\% of tasks due to diverse types
of failure. We used the same approach on Summit: eliminating RP waiting time
with PRRTE, lead to between 3 and 10\% task failure rate with 2 total
execution failures out of 8 runs.

Based on the analysis of the failures we recorded, we configured PRRTE to use
a flat communication hierarchy and \texttt{ssh} as its communication channel.
This reduced the internal performance of PRRTE and limited the total amount
of concurrent tasks that it can handle to \({\sim}20000\) but it also allowed
a more aggressive communication rate between RP and PRRTE\@. In Experiment 4,
we were able to reduce the waiting time from 0.1s to 0.01s and to use 4
concurrent sub-agents for RP\@. This increased the rate of communication
between RP and PRRTE both for each single sub-agent and globally, due to the
concurrency among sub-agents.

\begin{figure}
  \centering
  \includegraphics[trim=0 13 0 0,clip,width=0.49\textwidth]{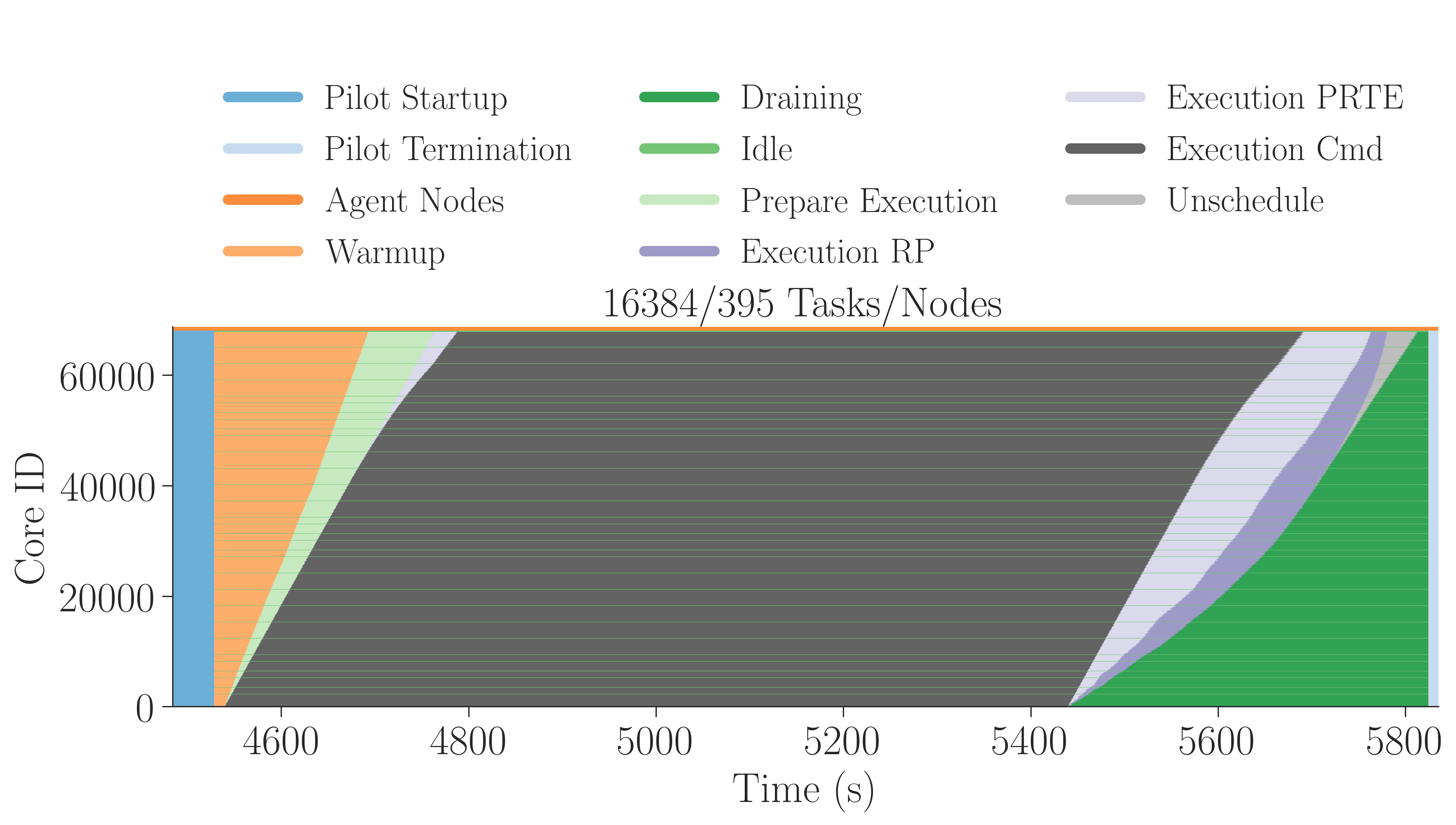}
  \Description{Area plot describing RP, PRRTE and workload
               resource utilization}
    \caption{\footnotesize Experiment 4: Resource utilization of a
             16384/404 tasks/nodes run with optimized RP/PRRTE
             integration.}\label{fig:ru_opt}
\end{figure}

Fig.~\ref{fig:ru_opt} shows how this dramatically improved RU\@. Compared to
the same run on Experiment 3, Experiment 4 reduced the mean of TTX from 3236s
to 1296s, the mean of aggregated RP overhead from 2648s to 522s, and the mean
of aggregated overhead of PRRTE from 2228s to 341s.

The last line of Table~\ref{tab:ru} shows the details of RU for Experiment 4.
RU of the workload improved from 25\% to 63\% while the RU of preparing
execution decreased from 29\% to 2\% and RU of idling resources decreased
from 8\% to 3\%. RU of both RP and PRRTE grows: RP deployment requires more
time due to the increased number of components instantiated and the higher
rate of scheduling; and the increased rate of task scheduling, placement and
execution stress more the capabilities of RP and PRRTE implementations.

Eliminating the delay between RP and PRRTE requires to introduce concurrency
between the two systems. We prototyped a version of RP that partitions the
available resources and uses a DVM for each partition. Tasks will be
scheduled across partitions, adding a minimal overhead for meta-scheduling
operations. Each DVM will operate on smaller portions of resources, lowering
the number of tasks scheduled by RP\@. In turn, this will eliminate the need
to add a waiting time, further reducing the overheads measured in our
baseline and improving both TTX and RU\@.

The main space for further improvement in RP is scheduling efficiency. RP
already support a scheduling framework in which algorithms tailored to the
executed workload can be used to optimize performance. Nonetheless, in RP the
scheduling algorithms are implemented in Python, leading to an unavoidable
ceiling on the amount of optimization we can implement at large scales.
Prototypes implemented in C show the near complete elimination of scheduling
overheads when executing both heterogeneous and homogeneous workloads.
Integration with third-party tools like Flux~\cite{ahn2014flux} is another
promising approach to solve this problem.

PRRTE overheads for individual tasks are both stable and small. This leaves
little space of optimization for the execution of many-tasks workloads as
those scheduled by RP\@. The main improvement for PRRTE would be to increase
the number of concurrent tasks that can be managed by the DVM but the
importance of such an improvement will decrease once RP can utilize multiple
DVMs within the same pilot.

% ---------------------------------------------------------------------------
% 2. RELATED WORK
% ---------------------------------------------------------------------------
\section{Related Work}\label{sec:related}

Pilot systems like GlideinWMS~\cite{sfiligoi2008glideinwms},
PanDA~\cite{de2014panda} and DIRAC~\cite{tsaregorodtsev2004dirac} are used to
implement late binding and multi-level scheduling on a variety of platforms.
While these systems have been successfully used on HPC
machines~\cite{hufnagel2017cms,maeno2014evolution,fifield2011integration},
including on the former ORNL leadership class machine
Titan~\cite{htchpc2017converging}, they are currently not available on Summit
and do not support either JSM or PRRTE\@.

Both PRRTE~\cite{castain2018pmix} and JSM~\cite{quintero2019ibm} rely on PMIx
to place and launch processes on Summit's nodes. Many applications are
actively working to directly use PMIx to interface with the system management
stack to benefit from portable process and resource management
capabilities~\cite{vallee2018improving}. While PMIx explicitly supports
interfacing with command line tools, there are no other pilot systems using
PMIx via JSM or PRRTE\@. MPICH and Hydra~\cite{balaji2014mpich} offer
capabilities similar to PRRTE but are not supported on Summit.

Pilot systems are not the only way to execute many-task applications on HPC
machines. JSM and LSF natively support this capability but, as seen
in~\S\ref{sec:performance}, in their current deployment on Summit they cannot
scale beyond 1000 concurrent task executions. Flux~\cite{ahn2014flux} is a
resource manager that provides users with private schedulers on pools of
dedicated resources. This enables the task scheduling capabilities of a pilot
system, including RP, but requires to be either adopted as the main job
manager of the machine or be deployed as part of a pilot system.

METAQ~\cite{metaq,metaq-2} are a set of shell scripts that forms a ``middle
layer'' between the batch scheduler and the user’s computational job scripts
and supports task packing. METAQ requires a separate invocation of mpirun (or
equivalent) for each task. METAQ has been superseded by
\texttt{mpi\_jm}~\cite{mpi-jm} --- a python library that is linked to
applications. In addition to intelligent backfilling and task packing,
\texttt{mpi\_jm} allows the executable to be launched based upon an affinity
with the hardware.

In Ref.\cite{cug-2016,merzky2018using} we investigated the performance of RP
on ORTE --- a precursor to PRRTE. Using ORTE, RP was capable of spawning more
than 100 tasks/second and the steady-state execution of up to 16K concurrent
tasks. Resource utilization was significant lower than with PRRTE and more
sensitive to the number of units and unit duration.

% ---------------------------------------------------------------------------
% 5. CONCLUSIONS
% ---------------------------------------------------------------------------
\section{Conclusions}\label{sec:conclusions}

We characterized the performance of the integration between RP, JSM and PRRTE
on Summit when executing many-task workloads. Our baseline characterizes
aggregated and individual overheads for each system, measuring resource
utilization for each available computing core. Our baseline measures the
performance for the worst case scenario in which single-core, 15 minutes-long
tasks are executed on up to 410 compute nodes of Summit. Further, based on
the insight gained by our characterization, we showed the results of our
optimization when executing 16384 1-core, 15 minutes-long tasks on up to 404
compute nodes.

Our experiments shows that on Summit: (1) PRRTE enables better scalability
than JSM when executing homogeneous many-task applications at scales larger
than 987 concurrent task executions; (2) up to the scale currently supported,
PRRTE individual overheads are negligible when compared to other overheads;
and (3) PRRTE open source code enables optimizations that lower the impact of
aggregated overheads over the execution of the considered workloads. Further,
we show that RP can effectively integrate with both JSM and PRRTE, imposing
manageable aggregated overheads while offering high degrees of
configurability. Specifically, we show that once optimized, at the largest
scale supported and for the considered workload the integration between RP
and PRRTE imposes an overall aggregated overhead of 35\% over the total time
to execution of the workload. This enables the utilization of 63\% of the
available resources to execute the given workload.

The presented performance characterization, its analysis, and the implemented
optimizations are the foundation of future work with RP, JSM, PRRTE and
Summit. The current scale at which RP/PRRTE operate support the development
of three use cases: machine learning driven molecular dynamics simulations;
machine learning driven drug discovery protocols; and seismic inversion
workflows. RP and PRRTE are posed to support several INCITE and Exascale
computing projects, accounting for a significant portion of the available
allocation on Summit in the next years. To this end, we will enable RP to
partition both available resource and workload execution. As seen
in~\S\ref{ssec:discussion}, this will greatly reduce aggregated overheads and
improve resource utilization efficiency. Further work will be needed to
optimize RP scheduler when managing workload with both spatial and temporal
heterogeneity, i.e., those in which task execution time are drawn from a
large distribution. The next step will be to characterize performance at
increasingly large scales, while measuring and addressing the bottlenecks for
heterogeneous workloads executed both concurrently and sequentially on the
same pilot.

% ---------------------------------------------------------------------------
% ACKNOWLEDGMENTS
% ---------------------------------------------------------------------------
\section{Acknowledgments}

We would like the thank other members of the PMIx community, and Ralph
Castain in particular, for the excellent work that we build upon. This
research used resources of the Oak Ridge Leadership Computing Facility at the
Oak Ridge National Laboratory, which is supported by the Office of Science of
the U.S. Department of Energy under Contract No.~DE-AC05-00OR22725. At
Rutgers, this work was also supported by NSF ``CAREER'' ACI-1253644,
RADICAL-Cybertools NSF 1440677 and 1931512, and DOE Award DE-SC0016280. We
also acknowledge DOE INCITE awards for allocations on Summit.

% ---------------------------------------------------------------------------
% BIBLIOGRAPHY
% ---------------------------------------------------------------------------
\bibliographystyle{acm}
\bibliography{prrte-jsrun}

\end{document}